\documentclass[wssquare]{ws-procs961x669} 
\usepackage{dcolumn}
\usepackage{bm}

\usepackage{graphicx}
\usepackage[tight,footnotesize]{subfigure}
\usepackage{color}
\usepackage[colorlinks,citecolor=blue]{hyperref}

\usepackage{multirow}%
\usepackage{float}
\usepackage{perpage}
\usepackage{morefloats}

\usepackage{latexsym}
\usepackage{gensymb} 
\usepackage{bbm}

\begin{document}
	
\title{String self-interactions near the deconfinment point in SU(3) Yang-Mills theory}

\author{A. Bakry$^*$, X. Chen$^\dag$, M. Deliyergiyev$^\ddag$, S. Xu$^\S$, P.M. Zhang$^\odot$}

\address{	
Department of High Energy Nuclear Physics,
Institute of Modern Physics,\\ 
Chinese Academy of Sciences, Lanzhou,  730000, China\\
$^\dag$E-mail: xchen@impcas.ac.cn\\
$^\ddag$E-mail: deliyergiyev@impcas.ac.cn\\
$^\S$E-mail: xsq234@163.com\\
$^\odot$E-mail: zhpm@impcas.ac.cn\\
http://english.imp.cas.cn}

\author{A. Galal and A. Khalaf$^\Diamond$}
\address{Physics Department, Faculty of Science, \\Al-Azhar University,
Cairo 11651, Egypt\\
$^\Diamond$E-mail: alikhalaf@alazhar.edu.eg}


\begin{abstract}
   We compare the predictions of Nambu-Goto effective string model at two loop-order for the Casimir energy and the width of the quantum delocalization of the string to the corresponding quark-antiquark potential and width profile of the color tube in pure $SU(3)$ Yang-Mills LGT at 4-dimensions near the deconfinement point.     
   Minor corrections are returned, when considering NLO terms for both the $Q\bar{Q}$ potential and broadening of the color tube, at the temperature $T/T_c=0.8$.       
   At a closer temperature to the critical point $T/T_{c}=0.9$, we found that the NLO contributions, from the expansion of the Nambu-Goto string, have a significant effect in improving the match to lattice data in the intermediate distances scales--before the string breaks in full QCD.       

   The string's self-interactions contributes in the suppressed broadening of the string's width and the non-curved/squeezed profile along the string in the intermediate distance.   
\end{abstract}


\bodymatter

\section{Introduction}

Precise lattice measurements of the $Q\bar{Q}$ potential in $SU(3)$ gauge model are in consistency with the L\"uscher subleading term for color source separation commencing from distance $R = 0.5$ fm~\cite{Luscher:2002qv} -- a typical distance scale where the effects of the intrinsic thickness of the flux tube at finite temperature, $1/T_c$~\cite{Caselle:2012rp}, diminishes and the effective description is expected to hold. 
The L\"uscher correction to the $Q\bar{Q}$ potential 
have unambiguously identified with unprecedented accuracy in many models~\cite{Juge:2002br,HariDass2008273,Caselle:2016mqu,caselle-2002,Pennanen:1997qm,Brandt:2016xsp}. In addition, the string model predicts a logarithmic broadening~\cite{Luscher:1980iy} for the width profile of the string delocalization. The observations of this effect have been reported in several lattice simulations corresponding to the different gauge groups~\cite{Caselle:1995fh,Bonati:2011nt,PhysRevD.27.2944,HASENBUSCH1994124,Caselle:2006dv,Bringoltz:2008nd,Athenodorou:2008cj,Juge:2002br,HariDass:2006pq,Giudice:2006hw,Luscher:2004ib,Pepe:2010na}.

However, in the intermediate distances and high temperatures this simplified picture of the free bosonic string, derived on the basis of the LO formulation of the Nambu-Goto(NG) action, does not provide a good description.
For instance, substantial deviations from the string behavior have been found for the lattice data corresponding to temperatures near the deconfinement point~\cite{PhysRevD.82.094503,Bakry:2010sp,Bakry:2011zz,Bakry:2012eq}. The region extends behind source separation distances at which LO string model predictions are valid at zero temperature regime. A comparison with the lattice Monte-Carlo data showed the validity of the LO approximation at separations larger than $R=0.9$ fm~\cite{PhysRevD.82.094503,Bakry:2010sp,Bakry:2012eq} for both the $Q\bar{Q}$ potential and flux-tube width profile. The authors of Ref.\cite{refId0} come to a similar conclusion  by considering the length of the Y-string between any two quarks in the baryon ~\cite{Bakry:2014gea,Bakry:2016uwt,Bakry:2011kga}.

The fact that free string theory poorly explains the lattice data in the intermediate distance and high temperatures has induced interest in the numerical experiments to verify the validity of higher order corrections to the NG action\cite{Caselle:2004jq,Caselle:2004er}. However, these higher order terms in the NG action are non-universal,  that is depend on the gauge model under consideration\cite{Caselle:2004jq, Giudice:2009di}. Even in the NG framework not all the orders of the power expansion are believed to give a good description of the correct behavior of the strings in the intermediate region.

This calls for a discussion concerning the distance and temperature scales for which the effective NG string description in the two loop order approximation is valid and which is the target of this report.

In this proceeding, we discuss the lattice data corresponding to the $Q\bar{Q}$ potential from $SU(3)$ Yang-Mills theory in four dimensions, and corresponding predictions of the LO and NLO Nambu-Goto potential~\cite{Caselle:2004er} in Sec.\ref{QQ_potential}. 
In Sec.\ref{String_wProfile}, the width profile of the density distribution is compare to the mean-square width of the string fluctuations in both approximations. 
The main conclusions are drawn in the last section.

\section{Quark-antiquark potential at LO and NLO}
\label{QQ_potential}

The linearly rising property of the confining potential induced the conjecture~\cite{Luscherfr} that the Yang-Mills (YM) vacuum admits the existance of an idealized one dimensional string object transmitting the strongly interacting forces between the color sources. 

The partition function of the NG model in the physical gauge\footnote{The physical gauge is required for the path integral  Eq.~\eqref{eq:PI} to be well defined with respect to Weyl and reparameterization invariance.} is a functional integral over all the worldsheet configurations swept by the string,
\begin{equation}
Z(R,T)= \int_{{\cal C}} [D\, X ] \,\exp(\,-S_{NG}( X )),
\label{eq:PI}
\end{equation}
where the NG action after gauge fixing reads
\begin{align}
S_{NG}[X]& = \sigma\frac{R}{T}+\frac{\sigma}{2}\int_{0}^{L_{T}}d\zeta_{1} \int_{0}^{R}d\zeta_{2} \left( \nabla X \right)^{2} + ...
\label{eq:GeneralForm_NGAction}
\end{align}
With the string collective variables $X^{\mu}=(\sigma^{\alpha}, X^{i}(\sigma))$, which allows one to embed the string worldsheet into the target space, where $\sigma^{\alpha}(\alpha=1,2)$ is the worldsheet coordinates, $X^{i}$ is the two-dimensional Goldstone bosons.
For the periodic boundary condition along the time direction with extend equals to the inverse temperature, $L_{T}=1/T$, the string collective variables can be defined as $X (\zeta_{1}=0,\zeta_{2})= X (\zeta_{1}=L_T,\zeta_{2})$. 
The Dirichlet boundary condition at the source position $X(\zeta_{1},\zeta_{2}=0)= X(\zeta_{1},\zeta_{2}=R)=0$.

The Casimir potential can be extracted from the string partition function Eq.\eqref{eq:PI} as
\begin{equation}
V(R,T)=-\dfrac{1}{T} \log\left(Z(R,T)\right)
\label{eq:Casimir}
\end{equation}
Solving the path integral of Eq.\eqref{eq:GeneralForm_NGAction} and using Eq.\eqref{eq:Casimir} with $\zeta$-regularization scheme~\cite{PhysRevD.27.2944} yields the model-independent static potential at the LO approximation
\begin{equation}
V_{\ell o}(R,T)= \sigma R+(D-2)T~{\rm{ln}} \eta \left(i\tau \right)+\mu(T),
\label{eq:QQpotential_LO}
\end{equation}
where $\mu(T)$ is a renormalization parameter, $\tau=\frac{L_{T}}{2R}$ is the modular parameter of the cylinder, $\eta$ is the Dedekind eta function defined as
$\eta(\tau)=q^{1/24} \prod_{n=1}^{\infty}(1-q^{n})$ with $q=e^{-\frac{\pi L_{T}}{R}}$.

The second model-independent correction to the Casimir effect has been derived in Ref.\cite{PhysRevD.27.2944} from the explicit calculation of the two-loop approximation as 
\begin{equation}
V_{n\ell o}(R,T)= V_{\ell o}(R,T) -T~{\rm{ln}} \left(1-\dfrac{(D-2)\pi^{2}T}{1152 \sigma_{o}R^{3}}\left[2 E_4(\tau) +(D-4)E_{2}^{2}(\tau)\right] \right),
\label{eq:QQpotential_NLO}
\end{equation} 
where $E$ is the Eisenstein series given by 
\begin{equation}
E_{2k}(\tau)=1+(-1)^{k}\dfrac{4k}{B_{k}} \sum^{\infty}_{n=1} \dfrac{n^{2k-1}q^{n}}{1-q^{n}}.
\end{equation} 
From Eq.\eqref{eq:QQpotential_NLO} one can derive the string tension as a function of the temperature \cite{Giudice:2009di} up to the NLO as
\begin{equation}
\sigma(T)=\sigma_{0}-\frac{\pi(D-2)}{6}T^{2}-\frac{\pi^{2}(D-2)^{2}}{72\sigma_{0}}T^{4}-\frac{\pi^{3}(D-2)^{3}}{432\sigma_{0}}T^{6}+{\mathcal{O}}(T^{7}).
\label{eq:stringTens_NNLO}
\end{equation} 
\begin{figure}[t]
\centering
\subfigure[$V_{Q\bar{Q}}$ at $T/T_c=0.8$]{
\includegraphics[scale=0.205]{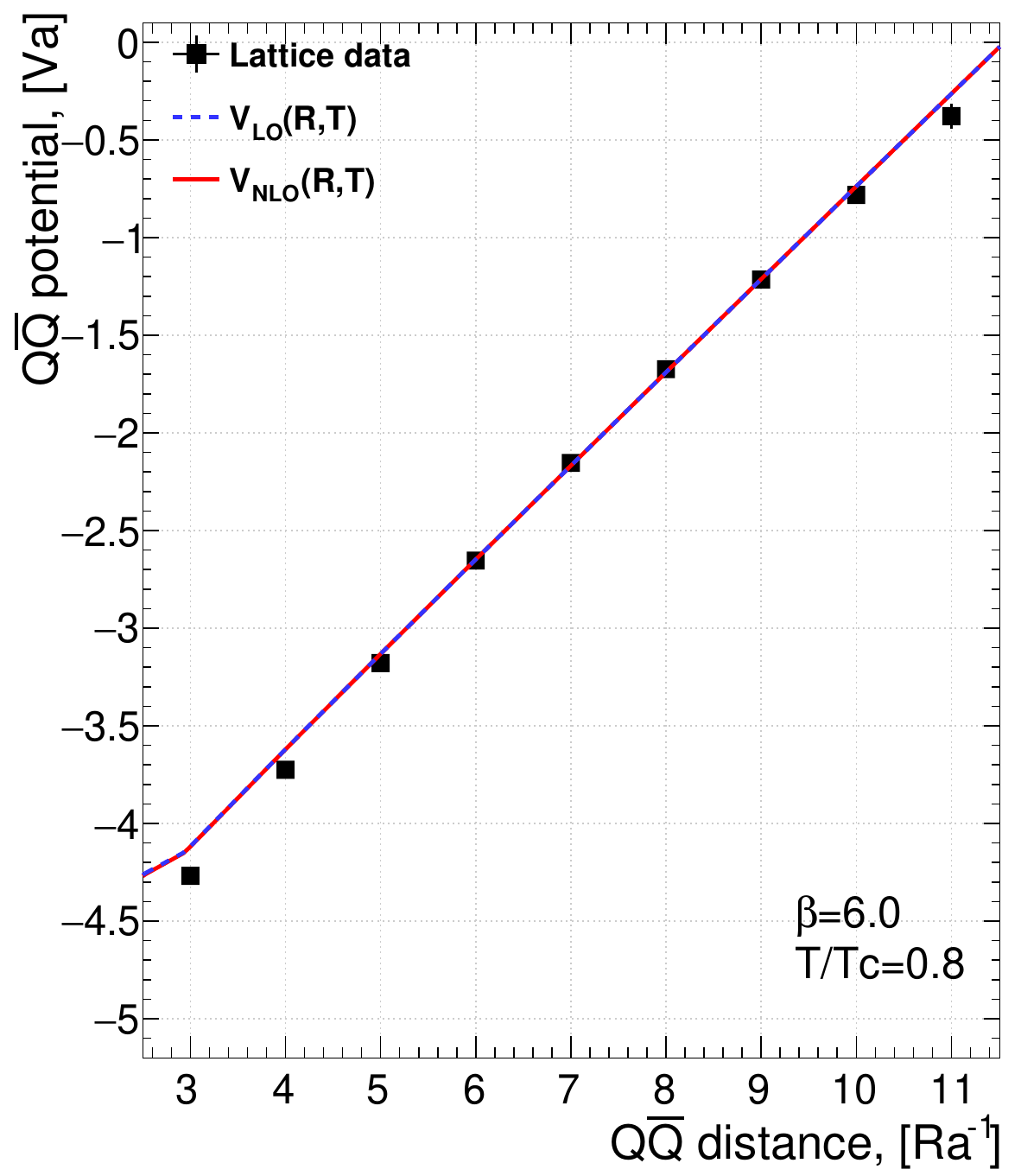}
}
\subfigure[Fit to $V_{Q\bar{Q}}$ using different fit ranges]{
\includegraphics[scale=0.40]{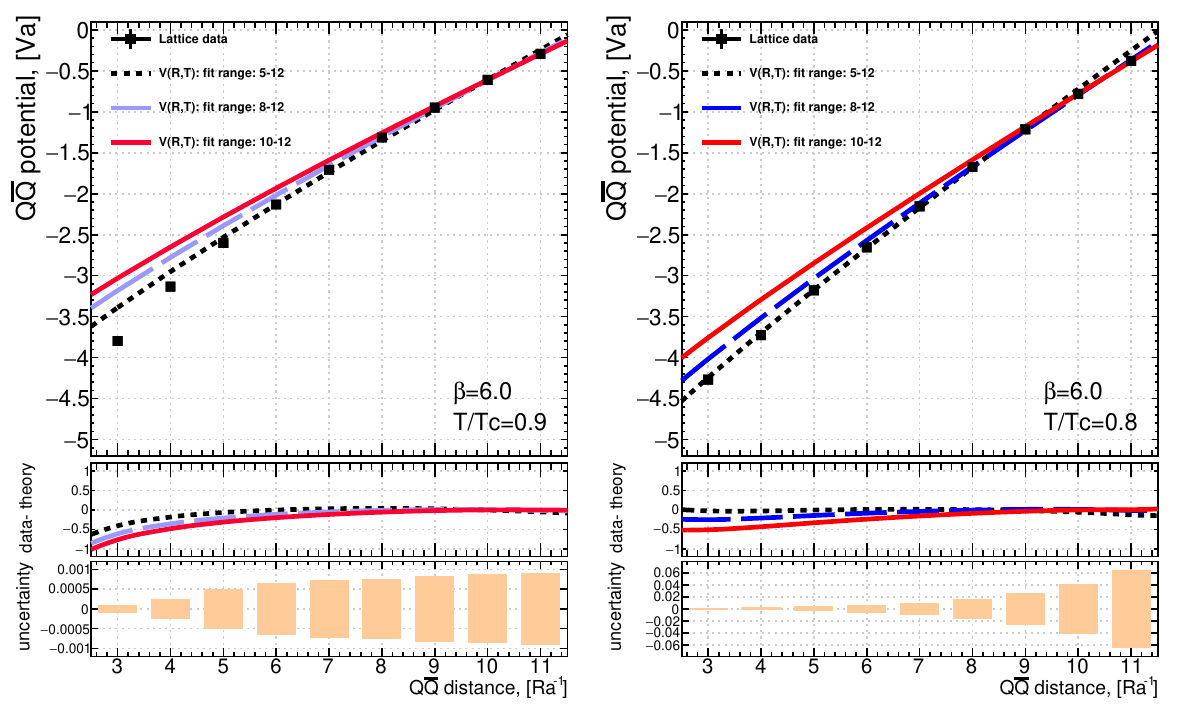}
}
\caption{
(a) The quark-antiquark potential, $V_{Q\bar{Q}}$, measured at temperature $T/T_c=0.8$, the solid and dashed lines correspond to fits to the LO and NLO string potential of Eq.\eqref{eq:QQpotential_LO} and Eq.~\eqref{eq:QQpotential_NLO}, respectively.
(b) $V_{Q\bar{Q}}$ measured at temperature $T/T_c=0.9$ (left), and at temperature $T/T_c=0.8$ (right), the lines correspond to the different fit ranges with a fixed end point at $R=12$ fm -- the different string tensions. Here we consider the string model solution only at NLO, see Eq.\eqref{eq:QQpotential_NLO}.
}
\label{fig:QQpot_08}
\end{figure}

The Monte-Carlo evaluation of the temperature-dependent ${Q\bar{Q}}$ potential 
\begin{equation}
V_{Q\bar{Q}}(R,T)=-\frac{1}{T} \log \langle P(x)P(x+R)\rangle
\label{eq:lattice_QQ_potential}
\end{equation}
at each $R$ is calculated through the Polyakov loop correlators 
\begin{equation}
\mathcal{P}_{\rm{2Q}} = \int d[U]P(0)P^{\dagger}(R) e^{-S_{w}} = e^{-\frac{V(R,T)}{T}},
\label{POT}
\end{equation}
with the single Polyakov loop given by 
$P(\vec{r}_{i}) = \frac{1}{3}\mbox{Tr} \left[ \prod^{N_{t}}_{n_{t=1}}U_{\mu=4}(\vec{r}_{i},n_{t}) \right]$.

The lattice data of the $V_{Q\bar{Q}}$, normalized to its value at $R=1.2$, at the temperature $T/T_c=0.8$, is shown on Fig.\ref{fig:QQpot_08}. To set comparison with the string model predictions, we fit the lattice results from Eq.\eqref{eq:lattice_QQ_potential} with the theoretical predictions of the NG string model at the LO and NLO separately, see Eqs.\eqref{eq:QQpotential_LO} and \eqref{eq:QQpotential_NLO}. 
The string tension $\sigma\, a^{-2}$ and the renormalization constant $\mu(T)$ 
was used as free parameters.

A large value of $\chi^2$ is returned for fits of color sources separations from $R=0.4$ to $R=1.1$ fm. For the separation distance $R\leq 0.4$ fm the NG string description is expected to show increasingly significant deviations from the lattice data due to the short distance physics and the idealized one dimensional NG string. 
Excluding the point $R=0.4$ fm from the fits significantly improved the data description for both the LO and the NLO approximations\cite{Bakry:2017utr}. We observed a manifesting stability in the returned values of the string tension parameter even by the exclusion of further points at short distances $R=0.5$ fm and $R=0.6$ fm from the fit range\cite{Bakry:2017utr}.  
The string tension settles a stable value of $\sigma a^{-2}=0.0445$ measured in lattice units at $T/T_{c}=0.8$, see Ref.\cite{Bakry:2017utr}.

The fit returns acceptable values of $\chi^2$ for the different fit ranges with a fixed end point at $R=12$ fm, as shown in Fig.\ref{fig:QQpot_08}(b). 
The values of the string tension depict a subtle corrections to the string tension compared to fits of the LO approximation of Eq.\eqref{eq:QQpotential_LO}. Within the standard deviations of the measurement, the zero temperature string tension stabilizes around the same value of $\sigma a^{-2}=0.046460$ for all data sets, for more details see Ref.\cite{Bakry:2017utr}. These results points out the minor role of the higher order modes due to the string's excited spectrum in the vicinity of the QCD plateau, $T/T_{c}=0.8$.

Thermal effects become more noticeable in the YM model~\cite{PhysRevD.85.077501,Doi2005559} considering the lattice with $N_{t}=8$ slices in the temporal direction, this scales the temperature to $T/T_{c}=0.9$ -- sufficiently close to the critical point. The lattice data corresponding to the measured  $V_{Q\bar{Q}}$ is depicted in Fig.\ref{fig:QQpot_08}(b-left). As a consistency check, we reproduced the same value of the string tension taken as a fit parameter as in Ref.\cite{PhysRevD.85.077501,Kac} with two different parameterizations of Ref.\cite{Gao,Luscher:2002qv} and fit domain. We follow different connive to scrutinize the fit behavior of the lattice data at this temperature scale with respect to the LO and NLO approximation, we schematically inspect the returned values of $\chi^{2}$ for an interval of selected values of the string tension $\sigma a^{-2} \in [0.036, 0.045]$. The residuals and normalization constant $\mu(T)$ for the corresponding $\sigma a^{-2}$ are listed in Ref.\cite{Bakry:2017utr}. 
The fits with the exclusion of points at short color distance separations reduces the value of $\chi^{2}$ for most of the selected values of the string tension. 

In spite of the fit result stretching out the role for the string's self interactions beyond the Gaussian approximation. The inclusion of the NLO terms, Eq.\eqref{eq:QQpotential_NLO}, provides a better $\chi^{2}$, limited to a color separation distance $R=0.5$ fm where indications for the validity of the string picture has been reported at zero temperature\cite{Luscher:2002qv}. The consideration of the two-loop approximation at $T/T_c=0.9$ finely corrects the free-parameter $\sigma a^{-2}$ interpreted as the zero temperature limit of the string tension for the value measured at $T/T_c=0.8$.

\section{The String Width Profile}
\label{String_wProfile}

In the following, we measure the mean-square width of the action density in $SU(3)$ gluonic configurations. The action density is related to the chromo-electromagnetic fields via $\frac{1}{2}(E^{2}-B^{2})$ and is measured through a three-loop improved lattice field-strength tensor~\cite{Bilson}. A color-averaged infinitely-heavy static $Q\bar{Q}$ state has been constructed by means of two Polyakov lines
$\mathcal{P}_{2Q}(\vec{r}_{1},\vec{r}_{2}) =  P(\vec{r}_{1})P^{\dagger}(\vec{r}_{2})$. 
A scalar field characterizing the action density distribution in the Polyakov vacuum or in the presence of color sources~\cite{Bissey} is defined as 
\begin{equation}
\mathcal{C}(\vec{\rho};\vec{r}_{1},\vec{r}_{2} )= \frac{\langle\mathcal{P}_{2Q}(\vec{r}_{1},\vec{r}_{2}) \, S(\vec{\rho})\,\rangle } {\langle\, \mathcal{P}_{2Q}(\vec{r}_{1},\vec{r}_{2})\,\rangle\, \,\langle S(\vec{\rho})\, \rangle},
\label{eq:Actiondensity}
\end{equation}
with the vector $\vec{\rho}$ referring to the spatial position of the energy probe with respect to some origin, and the brackets $\langle ... \rangle$ stands for averaging over gauge configurations and lattice symmetries. We make use of the symmetry of the four dimensional torus, that is, the measurements taken at a fixed color source's separations $R$ are repeated at each point of the three-dimensional torus and time slice then averaged. The lattice size is sufficiently large to avoid mirror effects or correlations from the other side of the lattice due to the periodicity of the discretized space-time mesh. Cluster decomposition of the operators leads to $C \rightarrow 1$ away from the quarks.

A measurement of the width of the string's action density may be taken by fitting the density distribution $\mathcal{C}(\vec{\rho}(r,\theta;z))$ from  Eq.~\eqref{eq:Actiondensity} to the following form
\begin{equation}
\mathcal{C}(\vec{\rho}(r,\theta;z))=1-G(r,\theta;z)=1-\left[A (e^{-r^2/\sigma_1^2}+e^{-r^2/\sigma_2^2})+\kappa \right],
\label{eq:width1}
\end{equation} 
taking into consideration the axial cylindrical symmetry of the tube, 
$r^2=x^2+y^2$ in each selected transverse plane $\vec{\rho}(r,\theta;z)$ to quark axis $z$.

The second moment of the action density distribution with respect to the cylinder's axis $z$ joining the two quarks is then given by
$W^{2}(z)=\frac{\int dr r^{3} G(r,\theta;z)}{\int dr r G(r,\theta;z)}$,
which defines the mean-square width of the tube. The localization of the color sources corresponds to $z=0$ or $z=R$, respectively.
\begin{figure*}[t]
	\begin{center}
		\subfigure{
			\includegraphics[scale=0.30]{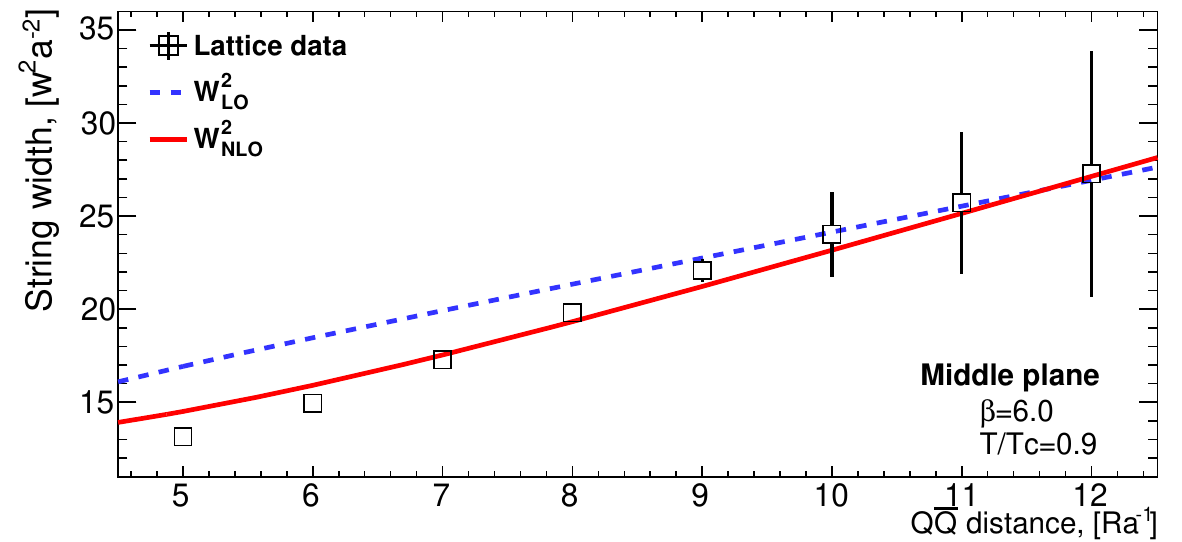}	
		}		
		\subfigure{
			\includegraphics[scale=0.30]{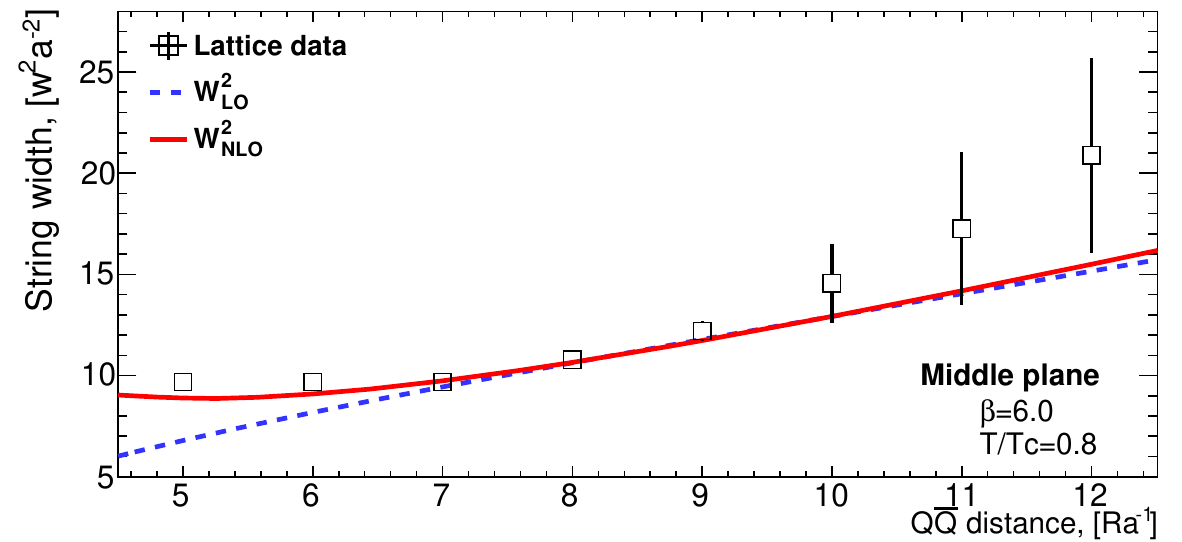}	
		}
		\caption{
			Plot of the mean-square width $W^{2}a^{-2}$ of the density distribution at the center of the tube, $z=R/2$, at temperature $T/T_c = 0.9$ (left) and $T/T_c = 0.8$ (right). The solid and dashed lines correspond to the free and self-interacting NG string Eq.\eqref{eq:WidthLO} and Eq.\eqref{eq:WidthNLO}.
		}
		\label{fig:MT08}	
	\end{center}
\end{figure*}

\begin{figure}[t]
\begin{center}
\subfigure{
\includegraphics[scale=0.30]{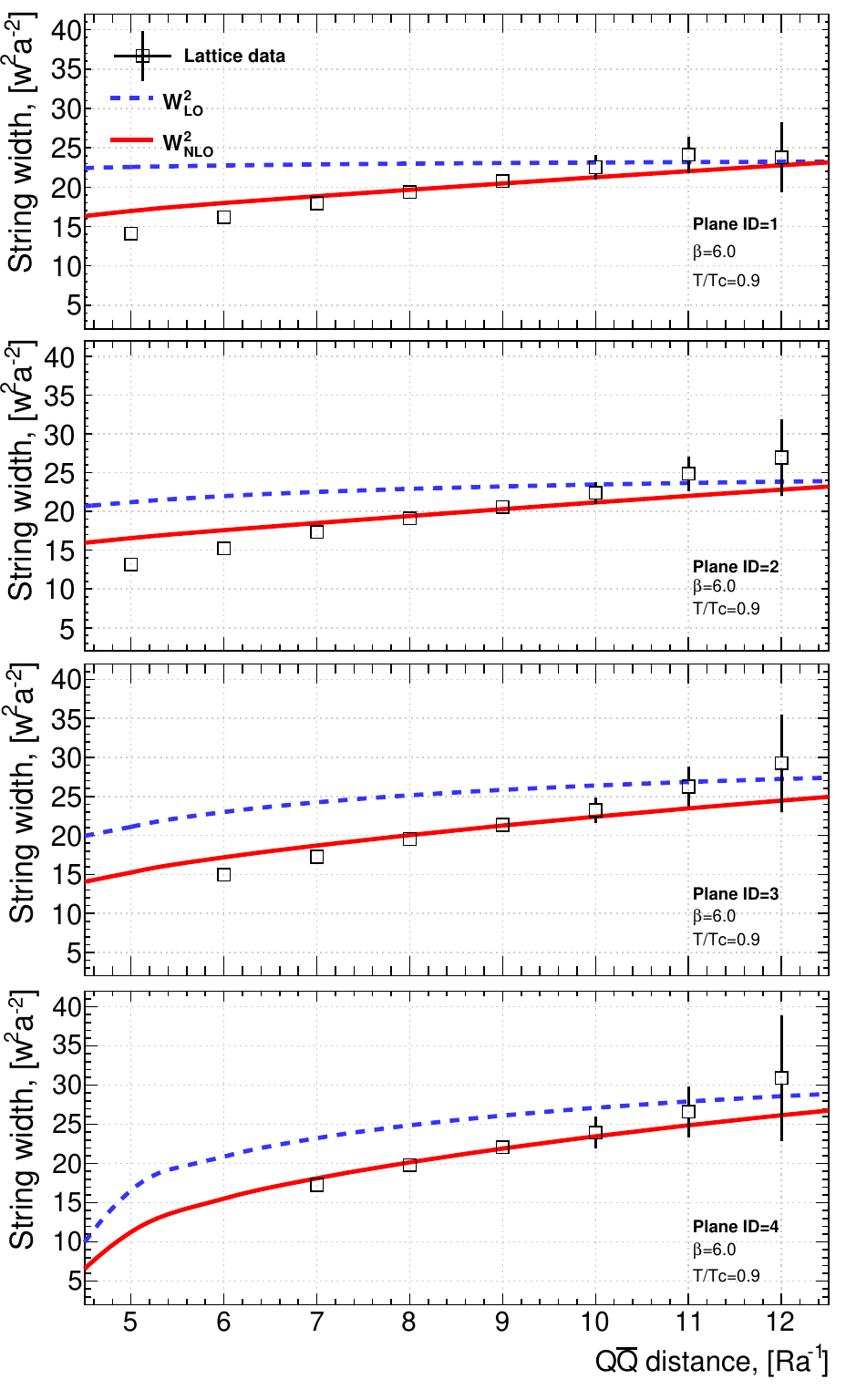}	
}						
\subfigure{
\includegraphics[scale=0.30]{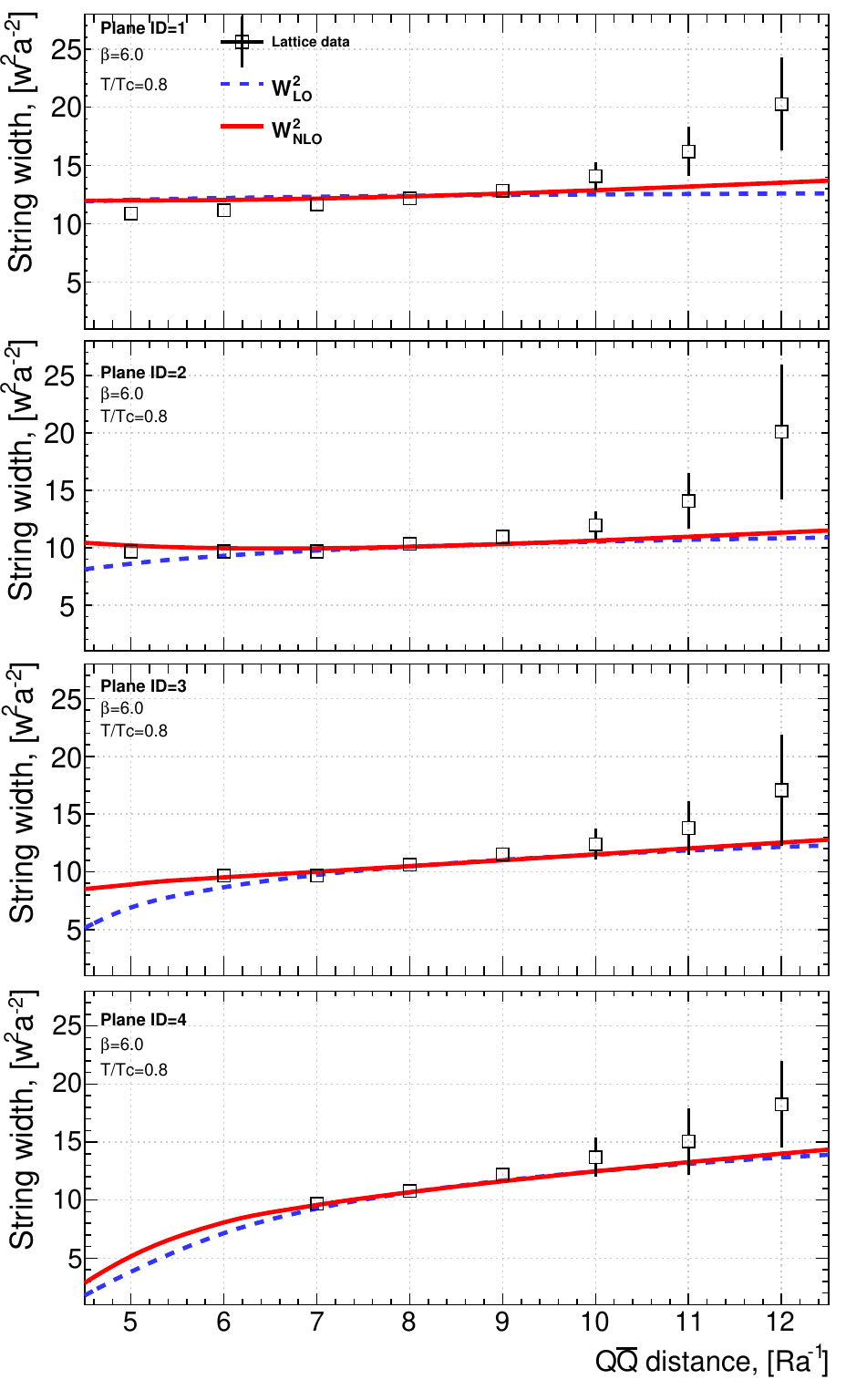}	
}
\caption{
The mean-square width of the string $W^{2}(z)$ versus quark-antiquark separations measured in the transverse planes at $z=1$, $z=2$, $z=3$, and $z=4$, from top to bottom. 
Measurements taken at temperature $T/T_{c}=0.9$ (left), $T/T_{c}=0.8$ (right). The dashed and solid lines denote the LO (Eq.\eqref{eq:QQpotential_LO}) and NLO (Eq.\eqref{eq:QQpotential_NLO}) one parameter string model, respectively.
}
\label{fig:PlanesID1234}						
\end{center}
\end{figure}

The width is estimated in accord to Eq.\eqref{eq:width1}
and $W^{2}(z)$ at each selected plane $z_i$ fixed with respect to one color source. 
The obtained values in middle plane $z=R/2$ are shown in Fig.\ref{fig:MT08}. Notice, that the mean-square width of the string at all middle planes exhibit a broadening as the color sources are pulled apart. The width at consecutive transverse planes $z_{i}$, see Fig.\ref{fig:PlanesID1234}, more clearly depict an increasing slop in the pattern of growth as one considers farther planes from the quark sources up to the middle plane.

The highest values of $\chi^{2}$ are retrieved if the whole source separations $R=4a$ to $R=12a$ are included for both LO and NLO approximations\cite{allais, Gliozzi:2010zv,Pepe:2010na}:
\begin{equation}
W^{2}_{{\ell o}}(\xi,\tau) = \frac{D-2}{2\pi\sigma}\log\left(\frac{R}{R_{0}(\xi)}\right)+\frac{D-2}{2\pi\sigma}\log\left| \,\dfrac{\theta_{2}(\pi\,\xi/R;\tau)} {\theta_{1}^{\prime}(0;\tau)} \right|,
\label{eq:WidthLO}
\end{equation}
\begin{equation}
\begin{split}
&~~~~~~~~~~~W^{2}_{n\ell o}(\xi,\tau)= \frac{\pi}{12 \sigma R^2}\left[E_2(i\tau)-4E_2(2i\tau)\right]\left(W_{lo}^2(\xi)-\frac{D-2}{4\pi \sigma}\right)\\
+&\frac{(D-2)\pi}{12\sigma^2 R^2}\Bigg\{\tau \left(q_{2} \frac{d}{dq_{2}}-\frac{D-2}{12}E_2(i\tau)\right)
\left[E_2(2i\tau)-E_2(i\tau)\right]-\frac{D-2}{8 \pi} E_2(i\tau)\Bigg\}.
\label{eq:WidthNLO}
\end{split}
\end{equation}
With the data points at short distances excluded from the fit the values of $\chi^{2}$ decrease gradually, indicating that only the data points at large source separation are parameterized by the string model formula. 

Most of the considerations on our discussion concerning the validity of both approximations to the ${Q\bar{Q}}$ potential at this temperature seem to hold for the string profile. 
The fits at $T/T_{c}=0.9$, however, entails that the LO approximation Eqs.\eqref{eq:WidthLO} are fitted on large distances to approach the NLO approximation in the asymptotic region.  
The increase of the uncertainties in the string's width  for color source separation $R\textgreater1.0$ fm does not loose our argument that the string model in both approximation scheme provides a good description for the string profile at this temperature. 

In spite of the improvements in the parameterization behavior at high temperatures $T/T_{c}=0.9$ in the intermediate distance region, the yet large values of $\chi^{2}$ raises the question whether it is sufficient to consider the NG action up to NLO or a more general string action encompassing the leading terms of NG action as a limiting approximation ought be considered. Nevertheless, the resolution of our lattice data is enough to disclose the roughness of the NG action expanded up to the fourth derivative terms in the precise description of the stringy color tube profile.

\section{Summary and Conclusion}

We consider the predictions of the NG string for the mean-square width profile near the end of the QCD plateau region, $T/T_{c}=0.8$. The obtained results indicate that the form of the string-vibration-like shapes of action density contours are independent from the temperature driven geometrical width effects suggesting that it can manifest even at lower temperatures. In addition, one may conclude that the potential extracted from the $Z(R,T)$ up to NLO are consistent with the lattice data for color source separations commencing from $R=0.5$ fm.

At the close enough temperature to the deconfinement point, $T/T_{c}=0.9$,  
the fits of the string model for either approximation schemes to the ${Q\bar{Q}}$ potential data return large values of $\chi^{2}$ considering a fit region spanning the whole intermediate region $R=0.5$ fm to $R=1.0$ fm. 
Fits considering both intermediate and asymptotic color source separation distances show significant improvement with respect to that obtained on the basis of the free string approximation. Nevertheless, we found that the NLO approximation does not provide an accurate match with the numerical data. 
This result is manifested in the returned values of $\chi^{2}$ which is still significantly large when considering distances less than $R\textless 0.8$ fm.

The NG effective description does not accurately describe the $V_{Q\bar{Q}}$ potential which manifest as a small deviation from the standard value of the string tension and the width profile of the quantum fluctuations of the string. This drives motives to scrutinize the fine structure of the strings up to the NNLO corrections or even higher order terms in the action with coefficients respecting the open-closed string duality~\cite{Luscher:2004ib}, or the general effective string actions of L\"uscher-Weiss action with boundary terms.

\section*{Acknowledgments}
This work has been funded by the Chinese Academy of Sciences President's International Fellowship Initiative grants No.2015PM062 and No. 2016PM043, the Recruitment Program of Foreign Experts, NSFC grants (Nos.~11035006,~11175215,~11175220) and the Hundred Talent Program of the Chinese Academy of Sciences (Y101020BR0). We thank Thomas Filk for helpful comments.

\bibliographystyle{ws-procs961x669}
\bibliography{SSISU3YM}

\end{document}